\documentclass[
    aps,
    pra,
    superscriptaddress,
    graphicx,
    amsmath,
    amssymb,
    reprint,
    floatfix,
    ]{revtex4-1}

\usepackage{graphicx} 
\usepackage{subcaption}
\usepackage{physics}
\usepackage{qcircuit}
\usepackage{xcolor}
\usepackage[normalem]{ulem}
\usepackage{verbatim}

\usepackage[colorlinks=true,citecolor=blue]{hyperref}

\newcommand{\uncertainty}[2]{\raisebox{0.5ex}{\tiny$^{+#1}_{-#2}$}}

\renewcommand{\sout}[1]{}
\newcommand{\Vikas}[1]{{\textcolor{black}{#1}}}
\newcommand{\Matt}[1]{{\textcolor{black}{#1}}}

\begin{document}

\setlength{\abovecaptionskip}{5pt plus 2pt minus 2pt}
\setlength{\textfloatsep}{5pt plus 2pt minus 2pt}


\title{Circuit-based leakage-to-erasure conversion in a neutral atom quantum processor} 



\author{Matthew N. H. Chow}
\email[]{mchow@unm.edu}
\thanks{These authors contributed equally to this work}
\affiliation{Sandia National Laboratories, Albuquerque, New Mexico 87123, USA}
\affiliation{Center for Quantum Information and Control (CQuIC), University of New Mexico, Albuquerque, NM 87131, USA}
\affiliation{Department of Physics and Astronomy, University of New Mexico, Albuquerque, NM 87131, USA}

\author{Vikas Buchemmavari}
\email[]{bsdvikas@unm.edu}
\thanks{These authors contributed equally to this work}
\affiliation{Center for Quantum Information and Control (CQuIC), University of New Mexico, Albuquerque, NM 87131, USA}
\affiliation{Department of Physics and Astronomy, University of New Mexico, Albuquerque, NM 87131, USA}

\author{Sivaprasad Omanakuttan}
\affiliation{Center for Quantum Information and Control (CQuIC), University of New Mexico, Albuquerque, NM 87131, USA}
\affiliation{Department of Physics and Astronomy, University of New Mexico, Albuquerque, NM 87131, USA}

\author{Bethany J. Little}
\affiliation{Sandia National Laboratories, Albuquerque, New Mexico 87123, USA}

\author{Saurabh Pandey}
\affiliation{Sandia National Laboratories, Albuquerque, New Mexico 87123, USA}

\author{Ivan H. Deutsch}
\affiliation{Center for Quantum Information and Control (CQuIC), University of New Mexico, Albuquerque, NM 87131, USA}
\affiliation{Department of Physics and Astronomy, University of New Mexico, Albuquerque, NM 87131, USA}

\author{Yuan-Yu Jau}
\affiliation{Sandia National Laboratories, Albuquerque, New Mexico 87123, USA}
\affiliation{Center for Quantum Information and Control (CQuIC), University of New Mexico, Albuquerque, NM 87131, USA}
\affiliation{Department of Physics and Astronomy, University of New Mexico, Albuquerque, NM 87131, USA}



\begin{abstract}
Atom loss errors are a major limitation of current state-of-the-art neutral-atom quantum computers and pose a significant challenge for scalable systems. 
In a quantum processor with cesium atoms, we demonstrate proof-of-principle circuit-based conversion of this form of leakage error to erasure errors via Leakage Detection Units (LDUs), which nondestructively map information about the presence or absence of the qubit onto the state of an ancilla. 
We benchmark the performance of the LDU using a three-outcome low-loss state detection method 
and find that the LDU detects atom-loss errors with $\approx$ 93.4\%  accuracy, limited by technical imperfections of our apparatus. 
We further compile and execute a SWAP LDU, wherein the roles of the original data atom and ancilla atom are exchanged under the action of the LDU, providing ``free refilling” of atoms in the case of
atom loss. This circuit-based leakage-to-erasure error conversion is a critical component of a neutral-atom quantum processor where the quantum information may significantly outlive the lifetime of any individual atom in the quantum register.
Finally, we demonstrate that LDUs may also be used to handle other forms of leakage errors where population moves to states outside of the computational subspace.
\end{abstract}


\maketitle 


\section{Introduction}

Neutral atoms have seen tremendous progress recently as a viable platform for quantum computation~\cite{Deutsch2000, Bluvstein2022, Endres6100, AtomComputing1000, BernienDualRydberg, Thompson10kTraps,KaufmanMidCircuit, ThompsonYb171, StrathclydeRB, BirklSupercharge, SchreckNarrowLine, MPQ-10k, Infleqtion2024}. The ease with which the system can be scaled to many qubits, combined with parallel gate control and mid-circuit reconfigurability result in a flexible platform, with strong potential for realizing fault tolerant quantum computation~\cite{Bluvstein2024, Cong_Lukin_QEC_Rydberg_PRX_2022, Wu_Puri_Thompson_2022_Nature_erasure, omanakuttan2024faulttolerant}. Leakage out of the computational subspace, however, especially by loss of the constituent atoms, is currently a substantial limitation in state-of-the-art systems \cite{Evered2023,ma2023high} and remains a significant obstacle to the long-term prospects. 
Some degree of atom loss is inevitable in any sufficiently long quantum circuit due to the shallow trap depth and finite vacuum lifetime. 
Transport of atoms for logic operations in the course of a circuit and trap shutoffs during entangling gates are also expected to result in atom heating and atom loss. 
Additionally, other forms of leakage to states outside of the computational subspace are expected from finite Rydberg state lifetime effects and off-resonant scattering from the trapping laser and possibly single-qubit lasers.
Leakage errors are difficult to correct in a fault tolerant way and generally require significant overhead~\cite{Aliferis_Terhal_2007_LRU,suchara2015leakage}. Even if leakage errors only constitute a modest fraction of total errors, they can reduce fault-tolerance thresholds to zero unless leakage reduction techniques are employed~\cite{suchara2015leakage}.

To deal with leakage errors~in neutral-atom platforms, protocols have been proposed~\cite{Cong_Lukin_QEC_Rydberg_PRX_2022, Wu_Puri_Thompson_2022_Nature_erasure, omanakuttan2024faulttolerant}  using hardware-specific methods~\cite{Wu_Puri_Thompson_2022_Nature_erasure, Omanakuttan_future} or circuit methods \cite{Cong_Lukin_QEC_Rydberg_PRX_2022}. 
In alkaline-earth-like atom encodings, some pathways that lead to leakage errors have been successfully detected by leveraging the particular level structure of these atoms~\cite{ma2023high,EndresErasureCooling}.
Detecting leakage errors converts them into an erasure errors, which are easier to handle than generic unknown errors \cite{Wu_Puri_Thompson_2022_Nature_erasure,Erasure_sensing}.
However, no experiment to-date has demonstrated coherence-preserving detection of atom-loss type leakage errors.

In the absence of such hardware-specific techniques, circuit-based methods, such as Leakage Detection Units (LDUs) and Leakage Reduction Units (LRUs), may be used to detect or reduce (without detecting) leakage errors~\cite{preskill1998fault,gottesman1997stabilizer}. 
LDUs map the bit of information specifying whether 
a data atom is within the qubit subspace or not onto the 0/1 states of an ancilla qubit, and do so without disturbing the information in the data qubit in the case of no leakage.  
These circuit-based approaches to leakage errors have been previously demonstrated in other hardware such as trapped ions and superconducting qubits \cite{QuantinuumRacetrack, Chen2016leakage, Stricker_Blatt_2020_loss, McEwen2021, Miao2023}.
While thresholds for common codes (e.g., the toric code) drop to zero is the presence of uncorrected leakage, finite thresholds exist when LDUs or LRUs are inserted~\cite{Aliferis_Terhal_2007_LRU, suchara2015leakage, BrownSurfaceCodeLeakage}.

\begin{figure*}
    \centering
    \begin{subfigure}{0.35\textwidth}
        \caption{\raggedright}
        \includegraphics[width=0.95\textwidth]{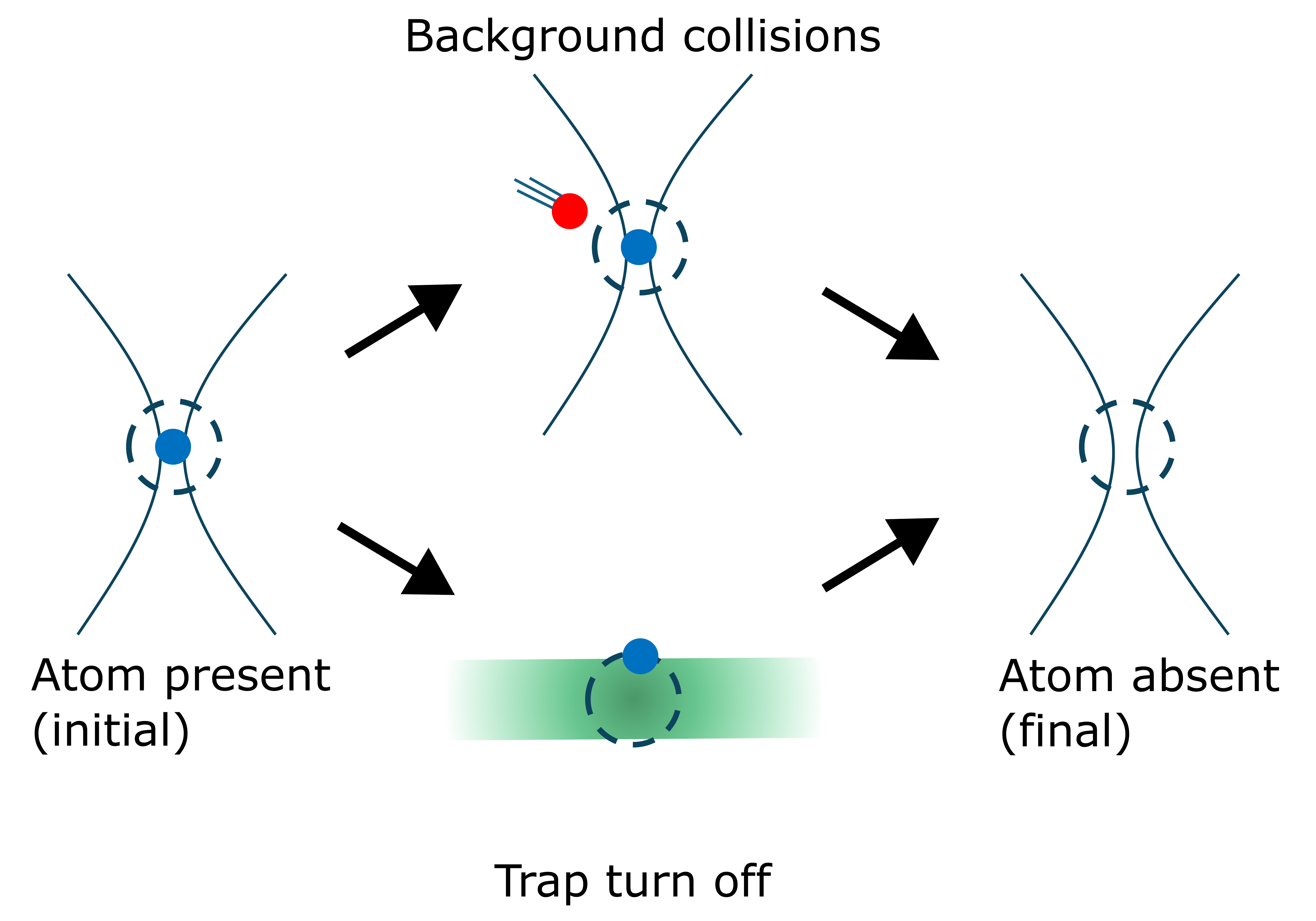}        
    \end{subfigure}
    \hspace{6pt}
    \begin{subfigure}{0.228\textwidth}
        \caption{\raggedright}
        \includegraphics[width=0.95\textwidth]{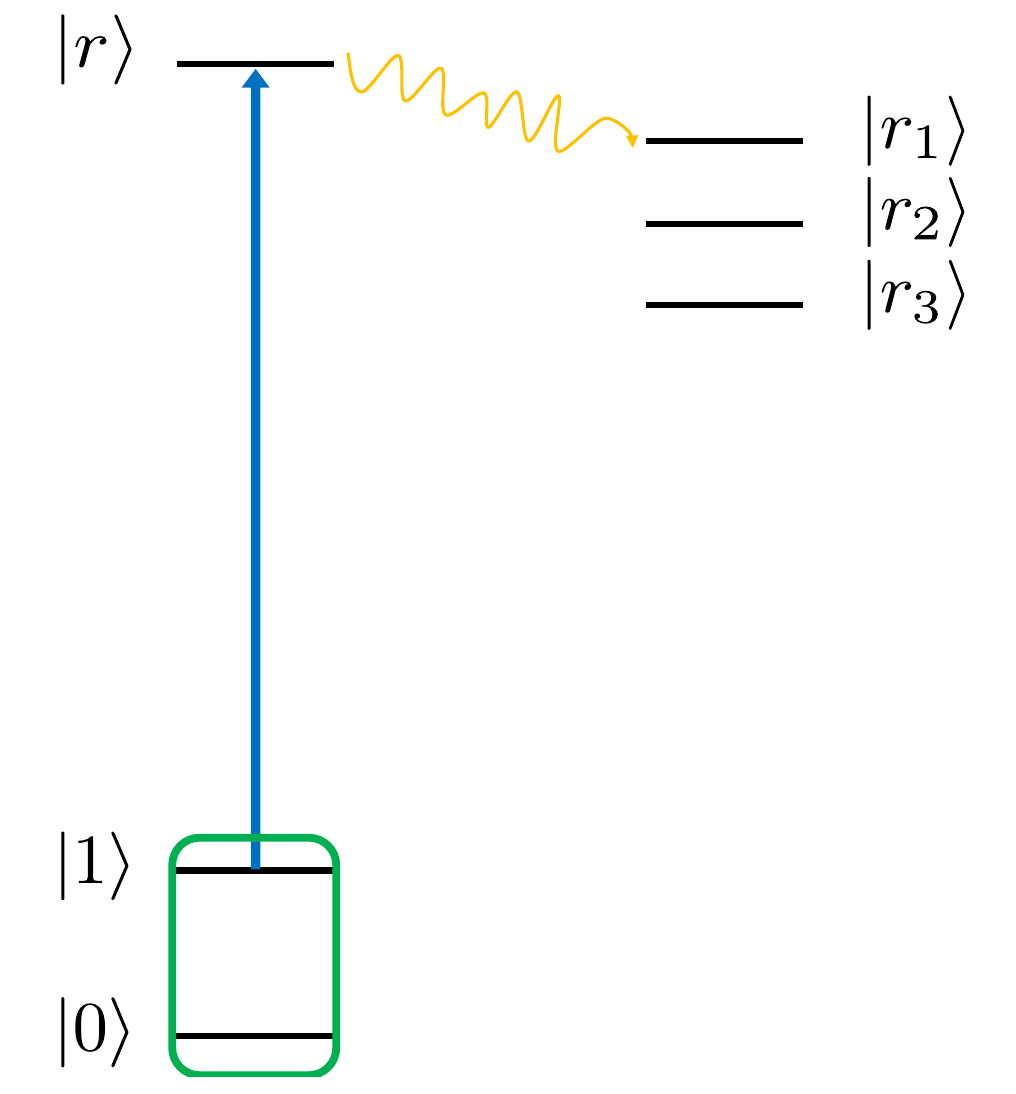}        
    \end{subfigure}
    \hspace{6pt}
    \begin{subfigure}{0.312\textwidth}
        \caption{\raggedright}
        \includegraphics[width=0.95\textwidth]{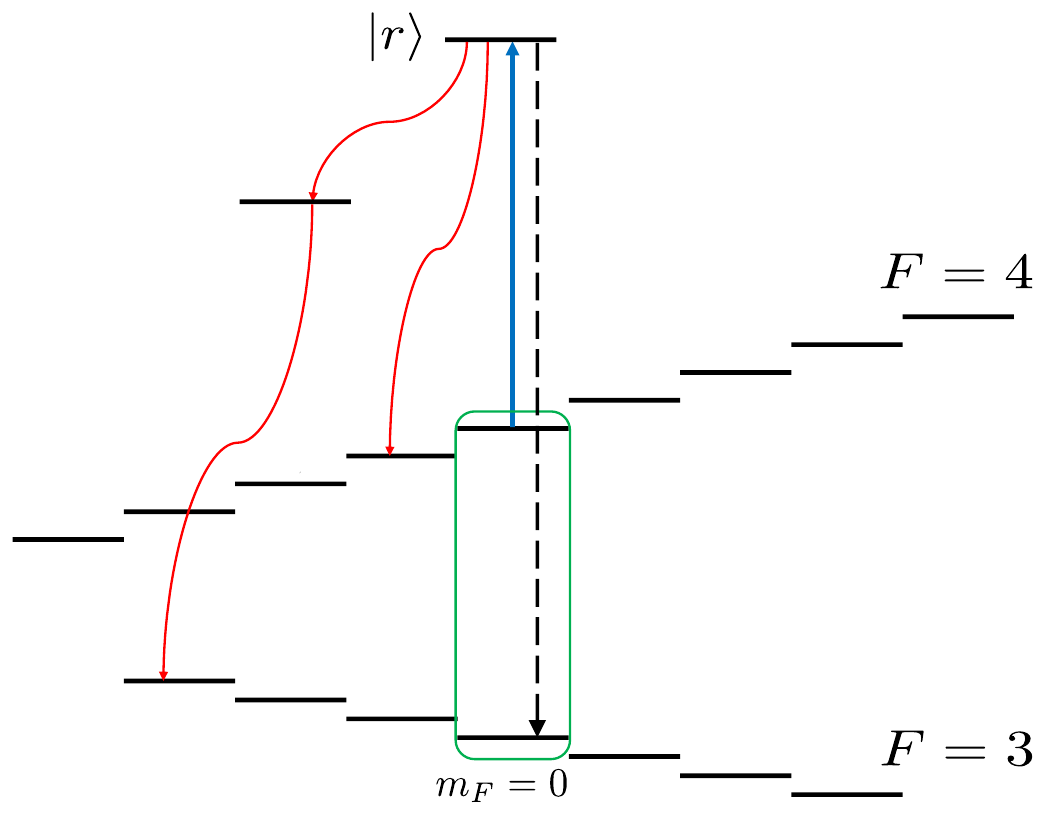}        
    \end{subfigure}
    \caption{\raggedright In addition to atom loss (a), we consider two other major leakage pathways (b, c) for Rydberg atoms considered in this work.
    \textbf{(a)} Atom loss from the tweezer may occur due to background gas collisions (as illustrated in the top pathway of (a)), heating (not shown), or imperfect recapture efficiency (as illustrated in the bottom pathway of (a)) after turning off the trap during Rydberg laser (green shaded region) operation.
    \textbf{(b)} During entangling gates, the population may be unintentionally left behind in the Rydberg state used for the gate ($\ket{r}$) and may decay to other Rydberg states ($\ket{r_{i}}$). This is called Rydberg leakage. 
    \textbf{(c)} Both Rydberg population decay and off-resonant scattering from Raman or trap lasers can cause so-called ``hyperfine leakage" (depicted as red curved arrows) into non-clock ground hyperfine sublevels, either directly or through intermediate excited states. Decay into the qubit subspace denoted by the green rectangle is not leakage, but instead falls under depolarizing or dephasing errors. 
    }
    \label{fig:leakage_paths}
\end{figure*}

In this work, we study the application of LDUs for neutral atoms and perform proof-of-principle demonstrations of their experimental implementation whereby we successfully detect leakage while preserving the coherence of the data qubit.
The objectives of an LDU are twofold: 1) detect whether or not some data atom is present and in the computational subspace, and 2) do so in a way that preserves the information in that data atom.  In this work, our qubit states are the ``clock-state" Zeeman sublevels in the hyperfine ground-state manifold of cesium atoms in optical tweezers. For hyperfine ground state qubits in alkali atoms, there are three major pathways for the atom to leak out of the computational space (see Fig.~\ref{fig:leakage_paths}). 
The first is atom loss from the trap due to background gas collisions, heating after many gates or movement operations, and  failure of recapture after the traps are turned off during entangling operations.  The second is Rydberg-leakage, where the population is unintentionally left behind in a Rydberg state after entangling operations, either due to coherent over-rotations or black-body induced transitions to nearby Rydberg states.  The third leakage mechanism is so-called hyperfine-leakage into non-qubit Zeeman sublevels of the hyperfine ground manifold due to decay from excited states.  Our objective is to demonstrate the detection of leakage via these three pathways with a circuit that preserves the coherence of the data atom. 

In Sec.~\ref{sec:standardldu} we describe the major leakage pathways in detail and show how they can be accounted for and detected using LDUs. This is achieved by turning Rydberg leakage into atom loss via the repulsive force of the trap for population left behind in a Rydberg state (Rydberg anti-trapping) and by using two-qubit gates that are sensitive to the choice of Zeeman sublevel.  We perform a proof-of-principle demonstration of the standard LDU, showing that the state of an ancilla atom correctly labels the presence or absence of a data atom within the qubit subspace without disturbing the information in the data atom.  Then in Sec.~\ref{sec:swapldu} we discuss how this LDU can be modified for neutral-atom platforms to provide ``free refilling'' of atoms by performing a SWAP.   We further implement a teleportation-based version of the SWAP LDU, which has only one entangling gate and free refilling capability. 

These demonstrations are made possible by a three-outcome measurement protocol (explained below in Sec.~\ref{sec:Methods}), which retains the atom in the trap after detection and can projectively detect atom loss by direct measurement, allowing us to benchmark the performance of the LDUs~\cite{BrowaeysDetection, HankinThesis, ChowDetection}. We also explore the benefits that three-outcome measurements confer upon LDUs.  In the absence of the three-outcome measurement capability, the circuits presented here may still be used to mitigate leakage errors, although in some cases they work only as Leakage Reduction Units (LRUs), which reduce, but do not detect leakage errors. 

In addition to detecting leakage errors, we show that the SWAP LDU provides a potential strategy for mid-circuit mitigation of atom heating and its resulting deleterious effects on atom loss and gate fidelity. For long circuits on this platform, heating of atoms will generally result from repeated applications of entangling gates (during which the optical tweezer trap is typically turned off and then atoms are recaptured) and from transport to rearrange the atomic configuration~\cite{ThomasTweezerHeating, SchulzThesis}.
As the SWAP LDU replaces a data atom with a ``fresh" cold ancilla atom, it is a powerful tool for maintaining a full, cold quantum register of data carriers by swapping in new atoms at a rate commensurate with atom loss and heating.

\subsection{Methods}\label{sec:Methods}
The entangling gate that we use throughout this work is a $ZZ$ rotation by angle $\theta$ via adiabatic Rydberg dressing \cite{KeatingCZ}. In all cases, we echo the single-atom light shift as in Refs.~\cite{Mitra2020, Martin2021} with a global single-qubit $\pi$-pulse between two Rydberg laser pulses. However, we use only the single-qubit $\pi$-pulse that performs the echo from that protocol, (dropping the outer $\frac{\pi}{2}$-pulses) so the three-pulse gate effectively performs the unitary $R_{zz}(\theta)R_{y}(\pi)$, see Fig.~\ref{fig:RzzCircuit}. 
The notation $R_{i}(\theta)$ is used throughout this work to represent rotations about Bloch-sphere axis $i \in \{x, y, z\}$ by angle $\theta$ according to the unitary, $\exp\left[-i\frac{\theta}{2}\hat{\sigma}_{i}\right]$, where $\hat{\sigma}_i$ is a Pauli operator. 
$R_{zz}(\theta)$ is the two-qubit analog representing the unitary, $\exp\left[-i\frac{\theta}{2}\left(\hat{\sigma}_z \otimes \hat{\sigma}_z\right)\right]$.
From Bell state fidelity measurements \cite{Sackett2000, Kim2009, ManningThesis, Figgatt2019} after 1, 5, and 9 applications of $R_{zz}(\frac{\pi}{2})R_{y}(\pi)$, we estimate the two-qubit gate 
fidelity to be 0.967(+5, -7) (see Fig.~\ref{fig:bellfidelity} and Appendix \ref{sec:Apdx_fidelity})
\footnote{For the spin echo of the entangling gate, we often choose to use a $\pi$-pulse about the $y$-axis to provide robustness to single-qubit pulse area errors when combined with preceding $x$-axis pulses \cite{CPMG}.}. 
Globally-addressed single-qubit $R_{\phi}(\theta)$ rotations by angle $\theta$ about equatorial axis $\phi$ are driven with a Raman beam as in Ref.~\cite{LukinDispersiveRaman} with short-timescale (not accounting for hours-scale experiment drift) fidelity $\approx 0.998(1)$ for a $\pi$-pulse,  estimated from the ratio of the Rabi rate and exponential decay rate of the Rabi oscillations. 
Globally-addressed single-qubit $R_z(\theta)$ gates are performed virtually via a phase advance of the local oscillator and are assumed to be error free. Individually-addressed single-qubit $R_z(\theta)$ operations are performed by using the trapping beams to induce a differential light shift \cite{MeschedeDifferentialLS}. These rotations are slow (tens of $\mu$s) and the fidelity of these operations is poor compared to the global single-qubit rotations, so they are used only when necessary (for individual state preparation and the teleportation experiment).

Independent state detection of each atom is performed via a low-loss state detection (LLSD) technique 
\footnote{Sometimes LLSD is referred to as nondestructive readout (NDRO) in literature. We use LLSD to avoid confusion over the term ``nondestructive'', which could accidentally imply non-projective measurements. LLSD is a projective measurement of both the atom state and the atom presence in the trap.} 
where (as in Fig.~2.8 of Ref.~\cite{HankinThesis}) the fluorescence from each trap site is coupled into an optical fiber and sent to a detector \cite{BrowaeysDetection, ChowDetection, HankinThesis}. Atoms are retained in the trap with high probability after state detection and we perform a second, state-insensitive detection stage to read out the final presence/absence of the atom in the trap. 
Therefore, LLSD is a three-outcome measurement revealing 0, 1, or ``neither'' (no atom), and we typically postselect against the neither condition.  The neither outcome is able to projectively measure atom loss with high accuracy and is used as a benchmark to measure the performance of the LDUs in this work.
The combined effects of state preparation and measurement (SPAM) errors and single-qubit gate errors from a $\pi$-pulse result in a measured single-atom SPAM fidelity of 0.990(7). 
Further details on the LLSD protocol are given in Appendix~\ref{sec:Apdx_LLSD}.
This $\approx 1\%$ SPAM uncertainty applies to all following results reported in this work (reported numbers for LDU performance are not SPAM-corrected). 
Further details of the apparatus are described in previous work \cite{ChowDetection}.

Finally, we note that during the theoretical analyses of LDUs, we neglect the errors caused by the LDUs themselves. We only consider the effects of ancilla atom loss errors that occur before the LDUs.

\section{Standard LDU With Two Entangling Gates}
\label{sec:standardldu}

We use an LDU adapted from~\cite{QuantinuumRacetrack,Stricker_Blatt_2020_loss,Cong_Lukin_QEC_Rydberg_PRX_2022} and originally proposed in \cite{preskill1998fault,gottesman1997stabilizer},
to nondestructively map a variety of leakage errors (including atom loss) in a data atom onto the state of an ancilla atom (see Fig.~\ref{fig:standardldu}). This ``standard" LDU, using the equivalent of two fully-entangling gates, is compiled to produce the identity on both the data ($d$) and ancilla ($a$) qubits ($I_d I_a$) in the case that no leakage occurred and to produce a bit flip on the ancilla ($X_a$) if leakage had occurred on the data atom prior to the LDU \footnote{Alternate compilations of LDUs may instead choose to perform the opposite logic and flip the interpretation of the readout result on the ancilla. We consistently use identity for no leakage and bit flip on the ancilla for convenience.}.
In the following sections we show that all three major leakage pathways for alkali atoms (atom loss, Rydberg leakage, and hyperfine leakage) can be detected using this standard LDU while preserving the coherence of the data atom. 

\subsection{Atom loss}
Since atom loss is the most dominant leakage channel in our system and is expected to be a critical challenge for neutral-atom experiments at scale, we begin by testing the LDU performance in detecting the presence of the data atom in the trap. We test the performance of the LDU against atom-loss errors by alternately preparing the data atom as being present or absent (in batches of 50). We directly verify the accuracy of the LDU against the final measurement of atom presence in the data atom site via LLSD. In the runs where LLSD reads out the data atom as present, the state of the ancilla atom correctly identifies atom presence in $765/814\approx 94 \%$ shots where the ancilla atom is also retained.
When the data atom is absent, the ancilla atom correctly identifies it in $919/984 \approx 93.4\%$ shots where the ancilla atom is retained. 
Inaccuracy of the LDU result when the data atom is present is dominated by gate errors (primarily two-qubit gates), whereas atom-absent inaccuracy is a combination of gate errors and finite probability of losing the data atom during or after the LDU. The measured atom-absent accuracy of the LDU is slightly better (865/901 $\approx 96\%$ of shots) in the case that the atom is intentionally prepared as absent, indicating better performance when a given atom-loss event is known to occur prior to the LDU (see appendix \ref{sec:shotbyshot} for full raw data).  As made possible by a three-outcome LLSD, these results are post-selected for ancilla atom retention. Post-selection resulting in exclusion of a shot of the experiment due to loss of the ancilla occurred in $202/2000 \approx 10\%$ of the total runs of the experiment. 

\begin{figure}
    \centering
    \begin{subfigure}{0.48\textwidth}
    \caption{\raggedright}
    $\begin{array}{c}
        \Qcircuit @C=.35em @R=.35em {
                & \lstick{\ket{\psi}_d} & \qw & \qw 
                & \ctrl{1} & \gate{X} & \ctrl{1} &  \gate{X} & \qw & \rstick{\ket{\psi}_d} \qw \\
                & \lstick{\ket{0}_a} & \gate{X} &  \gate{H} 
                & \ctrl{-1} & \qw & \ctrl{-1} & \gate{H} & \qw & \meter
                }
    \end{array}$
    \end{subfigure}
    
    \begin{subfigure}{0.48\textwidth}
        \caption{\raggedright}
        $\begin{array}{c}
        \Qcircuit @C=.35em @R=.35em {
                & & & & & & & \\
                & \lstick{\ket{\psi}_d} & \gate{{R}_{x}(\frac{\pi}{2})} 
                & \multigate{1}{R_{zz}(\frac{\pi}{2})R_y(\pi)} & \multigate{1}{R_{zz}(\frac{\pi}{2})R_y(\pi)} 
                &  \gate{{R}_{x}(\frac{\pi}{2})} & \gate{Z} \qw
                & \rstick{\ket{\psi}_d} \qw \\
                & \lstick{\ket{0}_a} &  \gate{{R}_{x}(\frac{\pi}{2})} 
                & \ghost{R_{zz}(\frac{\pi}{2})R_x(\pi)} & \ghost{R_{zz}(\frac{\pi}{2})R_y(\pi)} 
                & \gate{{R}_{x}(\frac{\pi}{2})} & \gate{Z} 
                & \meter 
                 \gategroup{1}{2}{2}{3}{.3em}{--}
                 \gategroup{1}{6}{2}{6}{.3em}{--}
                 \gategroup{3}{7}{3}{7}{.3em}{--}
                }
        \end{array}$
    \end{subfigure}

    \begin{subfigure}{0.48\textwidth}
        \caption{\raggedright}
        \label{fig:ramsey}
        \includegraphics[width=0.9\textwidth]{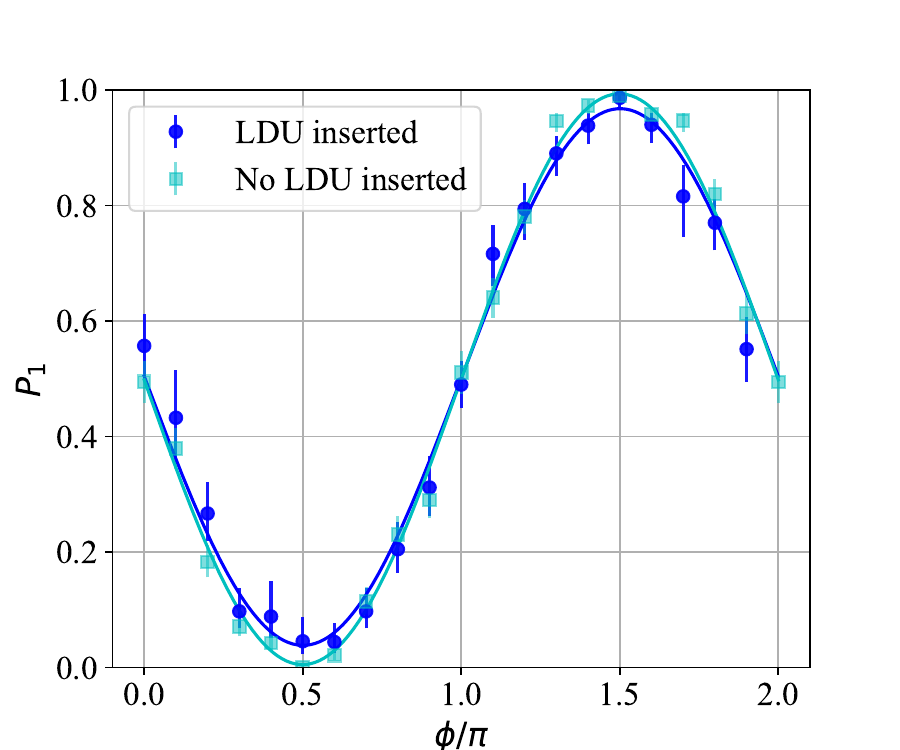}
    \end{subfigure}
    \caption{\raggedright 
    Standard leakage detection units (LDUs) using two entangling gates are composed to produce the identity on both the data ($d$) and ancilla ($a$) qubits ($I_d I_a$) in the case that no leakage occurred and produce a bit flip on the ancilla ($X_a$) if leakage had occurred prior to the LDU. 
    \textbf{(a)} The standard LDU composed of CZ, X, and H gates (typical native gates for neutral atoms) performs $I_{d}I_{a}$ in the case of no leakage and 
    $X_{a}$
    in the case that the data atom has leaked out the qubit subspace (wherein CZ becomes an identity). 
    \textbf{(b)} A standard LDU compiled for our native gates, using only globally-addressed single-qubit gates. The LDU implements $I_d I_a$ in the ideal case where no leakage has occurred and 
    $X_a Z_a$
    in the case where leakage has occurred. As a pair, the $R_{x}$ pulses on the data atom (enclosed in dashed boxes) are optional, and do not change the operation of the circuit. Similarly, as the final $Z$ gate on the ancilla is immediately followed by measurement, it is also optional.
    \textbf{(c)} A Ramsey coherence measurement on the data atom with (blue circles) and without (teal squares) insertion of the LDU as written in (b) demonstrates preservation of the quantum state of the data atom after the LDU. Probability of measuring the atom to be in the $\ket{1}$ state ($P_1$) is shown as a function of the phase ($\phi$) of a final $R_{\phi}(\frac{\pi}{2})$ pulse. Uncertainty markers on datapoints are 68\% Wilson score intervals.
    }
    \label{fig:standardldu}
\end{figure}

We also demonstrate that the coherence of the data atom is preserved after the LDU through a Ramsey experiment with the LDU inserted; see Fig.~\ref{fig:ramsey}. Without the LDU, contrast as determined by maximum likelihood estimation accounting for the underlying binomial distribution \cite{EndresErasureCooling} in the Ramsey experiment is 
99\Matt{(-2, +1)}\%, and with the LDU inserted, the contrast is 
93\Matt{(3)}\% \footnote{Uncertainty reported here is where the likelihood drops by a factor of 2 relative to the maximum (i.e., half width half max) in a one-dimensional cut along the contrast parameter, holding the other fitting parameters fixed at the optimum.}. The contrast reduction is consistent with our single- and two-qubit gate fidelities. As the entangling gate infidelity is estimated at 3.3(+7, -5)\% (see Fig.~\ref{fig:bellfidelity}), the two entangling gates of the standard LDU constitute the majority of the error that leads to loss in contrast. 

To confirm that the LDU is independent of input state of the data atom, we prepare the data atom in six different input states.  
As our desired circuit begins with $R_{x}(\frac{\pi}{2})$ on the ancilla atom and we wish to prepare various input states of the data atom, $\ket{\psi}_d$,
we prepare the data atom along the $x$-, $y$-, or $z$-axis of the Bloch sphere without impacting the state of the ancilla by following an initial global $R_{x}(\frac{\pi}{2})$ with a rotation by $R_z(\xi)$ about the $z$-axis on just the data atom and then a rotation $R_y(\theta)$ about the $y$-axis globally. Results for all six input states are given in Table~\ref{tab:2x2q} and confirm that the protocol is agnostic to the input state on the data atom.

\begin{table}
    \begin{tabular}{|c|c|c|c|}
        \hline
         $\ket{\psi}_{d}$ & Atom present accuracy & Atom absent accuracy \\
        \hline
        -y & 0.940(\uncertainty{7}{8}) & 0.934(\uncertainty{7}{8}) \\
        +y & 0.94(1) & 0.93(1) \\
        +x & 0.95(1) & 0.92(1) \\
        0 & 0.94(1) & 0.95(1) \\
        1 & 0.92(1) & 0.94(1) \\
        -x & 0.93(1) & 0.91(1) \\
        \hline
        Average & 0.938(4) & 0.931(4) \\
        \hline

    \end{tabular}
     \caption{\raggedright 
     Standard leakage detection unit performance with varied input state shows that the LDU works for all data atom input states. The accuracy of the LDU in labelling the data atom presence is verified by the state-independent detection stage of the LLSD. When the data atom is present (absent) as read out by the LLSD, the atom present (absent) accuracy displayed in column 2 (column 3) is assessed as the ratio of shots where the ancilla state after the LDU correctly labels the data atom presence over the total number of shots where the ancilla is retained. Uncertainties reported are 68\% Wilson score intervals, reported symmetrically in cases were upper and lower intervals have the same most-significant digit.
     }
     \label{tab:2x2q}
\end{table}

\subsection{Rydberg leakage}
Having demonstrated detection of atom presence with the standard LDU, we now turn to Rydberg state population leakage. 
Data atom population remaining in Rydberg states can still participate in entangling interactions with the ancilla (unlike atom loss) and lead to undesired outcomes for the LDUs. For example, if the data atom starts off in the Rydberg state used for the entangling gate, $\ket{r}$, during a CZ gate~\cite{KeatingCZ,LP_gate_2019,Jandura_Pupillo_gate_Quantum_2022,Pagano_PRR_2022_gate}, the ancilla atom ends up in $\ket{r}$ with finite probability, propagating the leakage error. For the Levine-Pichler  gate~\cite{LP_gate_2019,Jandura_Pupillo_gate_Quantum_2022,Pagano_PRR_2022_gate} and the adiabatic dressing CZ-gate~\cite{KeatingCZ}, if the two-atom initial state is $\ket{r1}$, the end state will be $\approx \alpha\ket{r1}+\beta\ket{1r}$ in addition to a possible phase error. If the initial state is $\ket{r'1}$, where $\ket{r'}$ is a different nearby Rydberg state populated due to blackbody-radiation, then only a coherent phase error is added on the second atom\Vikas{~\footnote{Here we assume that the state $\ket{r'}$ is transparent to the Rydberg laser being used but still fully blockades the state $\ket{r}$. In this case, the second atom does not undergo evolution during the entangling gate pulse due to the blockade, but the following one qubit phase cancellation introduces a phase error. If instead $\ket{r'}$ does not interact at all with $\ket{r}$, it falls in the category of less harmful leakage errors, like atom loss. In the intermediate case of imperfect blockade, the second atom can end in the Rydberg state, propagating the leakage error.} (also see concurrent work~\cite{jandura2024surface})}. In our experiment, if we initialize one atom in the Rydberg state ($\ket{r}$), we measure with $\approx$ 10\% probability that the other atom ends up in the Rydberg state after subsequent application of our $R_{zz}R_y$ entangling gate. 
This interacting form of leakage \cite{BrownSurfaceCodeLeakage} does not satisfy the ``sealed two-qubit gate" requirement~\cite{suchara2015leakage} and can propagate during entangling interactions.

Since the Rydberg  $64P_{3/2}$ state used in this work is anti-trapped for our trap wavelength (937\,nm), Rydberg state population errors may be naturally converted to atom-loss errors simply by turning on the trap \cite{Cong_Lukin_QEC_Rydberg_PRX_2022,Hines_MSS_spin_squeezing_PRL_2023}. 
We thus convert interacting Rydberg leakage into noninteracting atom-loss leakage, which is detectable with the LDU. 
We measure the speed at which a Rydberg atom is ejected from the trap by promoting an atom to the Rydberg state with a $\pi$-pulse, holding it for a variable amount of time, de-exciting it with a second $\pi$-pulse, and measuring the probability that the atom remains in the trap afterwards (see Fig.~\ref{fig:antitrapping}). In our system, the survival drops by $1/e$ in $\approx 23\,\mu$s, which is short compared to the Rydberg lifetime of $\approx 170\,\mu$s for the 64$P$ state at room temperature \cite{Martin2021, RydbergErrors2018}.
Trap wavelength, strength, geometry, and chosen Rydberg level all make the anti-trapping speed system-specific \cite{KokkelmansAntitrap}. 
Other, typically faster, methods of converting Rydberg population to atom loss utilize laser- or field-induced ionization \cite{Madjarov2020, ma2023high, Cong_Lukin_QEC_Rydberg_PRX_2022}.
Once converted to atom loss, Rydberg state population errors are detectable with the LDU as before. 
Alternate proposed methods provide for fast conversion of Rydberg leakage to Pauli errors \footnote{Any non-leakage error can be written as a linear combination of  Paulis, which means all non-leakage errors are effectively Pauli errors for error correction purposes.} by pumping the population to a short-lived intermediate state \cite{Cong_Lukin_QEC_Rydberg_PRX_2022,Hayes_Quantinuum_repumping_PRL_2020} or gate designs that prevent Rydberg leakage errors from spreading~\cite{jandura2024surface}.

With LLSD capability we typically postselect against atom-loss errors (from Rydberg anti-trapping or otherwise). Postselection on atom retention has been used to improve effective entangling gate fidelity at the expense of postselection rate \cite{Noel_infleqtion}. 
The entangling gate performance reported in this work similarly benefits from postselection against atom loss (and therefore Rydberg leakage converted to loss by anti-trapping), but we leave an in-depth study of this contribution to future work.  Our techniques are also compatible with recent developments from complementary work considering and mitigating non-leakage errors induced by unwanted Rydberg population \cite{jandura2024surface}.
The degree to which remaining Rydberg population contributes to the overall infidelity of two-qubit gates depends on the particular gate protocol used and has been studied previously \cite{Evered2023, PritchardBenchmarking, Infleqtion2024}. 


\subsection{Hyperfine leakage}

Finally, we test the performance of this LDU against hyperfine-leakage (leakage errors to other Zeeman levels in the ground state). We operate our experiment at $3.89\,$G bias magnetic field such that neighboring sublevels in the ground (Rydberg) manifold are spectrally separated by $1.36\,$MHz ($7.2\,$MHz). Therefore, neighboring sublevels are not expected to couple strongly to either the single- or two-qubit drive lasers and will be detectable as leakage errors via the LDU.

Our qubit computational basis states are the $\ket{F=4, m_F=0} \equiv \ket{1}$ and $\ket{F=3, m_F=0} \equiv \ket{0}$ states in the $6S_{1/2}$ electronic ground state manifold in Cs.  By pulsing on an additional, partially-overlapped tweezer (generated with an additional radio-frequency tone applied to the acousto-optic modulator) at a frequency separation equal to the Zeeman splitting (1.36\,MHz), we are able to locally induce coupling between neighboring Zeeman levels in the ground state manifold via stimulated Raman transitions. Although the trapping light is nominally $\pi$-polarized, which should not permit transitions to $\Delta m_F \neq 0$ states, the offset in position in the trapping plane and polarization vortices on the sides of the tightly-focused beams \cite{LukinSBC} provide sufficient polarization components to generate leakage to $m_F \neq 0$ levels. Using a 40\,$\mu$s pulse of the additional tweezer, we drive leakage out of the qubit level with $\approx$ 95.5\% (95.7\%) probability after preparing in the $\ket{0} (\ket{1})$ state. As the leakage-inducing pulse simply couples all neighboring Zeeman sublevels, the few percent remaining population in the clock state after the leakage pulse is expected.

When we use this post-leakage-inducing pulse state (a superposition of many $m_F$ levels within one $F$ manifold) as the data atom input state to the LDU, the ancilla reads out that the atom has leaked in 371/415 $\approx 89\%$  (355/407 $\approx$ 87\%) shots of the experiment when preparing in a mixture of $F=3$ $(F=4)$ Zeeman sublevels. 
We emphasize that these outcomes are consistent with (within shot-noise uncertainty of) the ideal case where the LDU would behave the same for non-qubit Zeeman sublevels as it would for an absent atom. 
Given the probability of remaining within the qubit subspace after the leakage-inducing pulse and atom-present accuracy of the LDU for the qubit states, the expected outcomes in the ideal case would be 90\% (88\%) leakage identification for the $F=3$ ($F=4$) experiment.
Any small remaining discrepancies in performance between LDU identification of Hyperfine leakage and atom-loss leakage (especially for the $F=4$ non-qubit states) may be due to weak, off-resonant coupling to the Rydberg state. We leave this topic as a subject of further study which may be experimentally probed using shelving schemes~\cite{Graham_Saffman_mid_circuit,Buchemmavari_Spin_flip_2024_PRA}. We note that operation at stronger bias magnetic fields may be required to achieve better Zeeman-state selectivity if the Rabi rate of the Rydberg laser is increased (as would be desirable for better entangling gate performance).

\section{SWAP LDUs}
\label{sec:swapldu}

\begin{figure*}
    \centering
    \begin{tabular}{c c}
        \begin{subfigure}{0.48\textwidth}
        \caption{\raggedright}
        $\begin{array}{c}
             \Qcircuit @C=1em @R=.7em {
                 \lstick{\ket{\psi}_d} & \ctrl{1} & \gate{X} & \ctrl{1} & \gate{X} & \qw & \ket{\psi}_d \\
                  \lstick{\ket{0}_a} & \targ & \qw & \targ & \gate{X} & \meter &  \\
                 & & & &
                }
        \end{array}$
        \end{subfigure}
     &  
         \begin{subfigure}{0.48\textwidth}
            \caption{\raggedright}
            \label{fig:refilling-swap}
            $\begin{array}{c}
                \Qcircuit @C=1em @R=.7em {
                 \lstick{\ket{\psi}_d} & \ctrl{1} & \targ & \meter &  \\
                 \lstick{\ket{0}_a}    & \targ    & \ctrl{-1}    & \qw & \ket{\psi}_a \\
                 & & & &
                }
            \end{array}$
        \end{subfigure}
     \\
    \begin{subfigure}{0.45\textwidth}
        \includegraphics[width=0.9\textwidth]{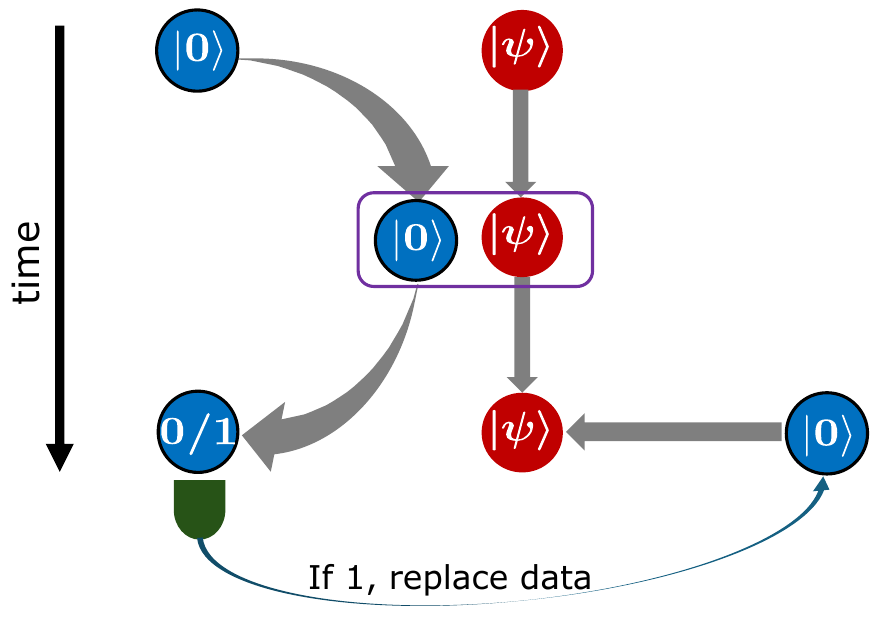}
    \end{subfigure}
     & 
    \begin{subfigure}{0.45\textwidth}
        \includegraphics[width=0.9\textwidth]{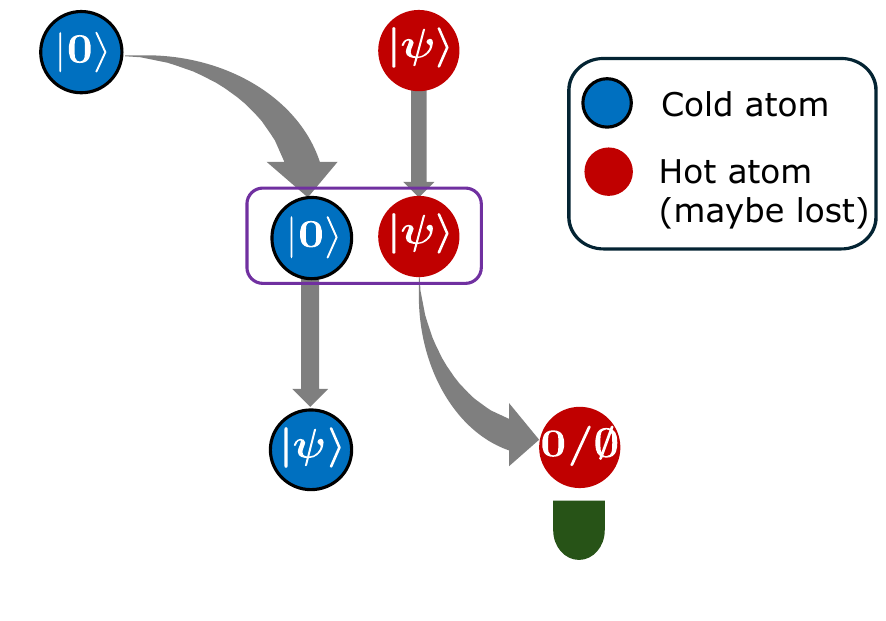}
    \end{subfigure}
    \end{tabular}
    \caption{\raggedright Refilling advantages of SWAP LDUs. The data atom (red, no outline) carrying state $\ket{\psi}$ is hot after going through many rounds of gates and rearrangement, and might have been lost or leaked. A fresh, cold ancilla atom (blue, dark outline) carrying $\ket{0}$ is brought close to the data atom to apply an LDU. 
    \textbf{(a)} In the standard LDU, the ancilla carries the information about the presence/absence of data atom and needs to moved away for measurement. If the measurement gives $1$, the data atom would need to be replaced with another fresh ancilla. 
    \textbf{(b)} In the SWAP LDU, the LDU transfers the data $\ket{\psi}$ onto the cold ancilla atom. The old data atom is borne away for a measurement. If the data atom was lost, it has already been replaced by the cold ancilla, removing the need for the extra refilling step. 
    In addition, the data is always transferred onto a fresh cold atom. This kind of cooling by replacement may reduce potential problems from increasing atom temperature in operations later in the circuit.
    }
    \label{fig:refilling}
\end{figure*}

For neutral-atom platforms, there may be further advantages of LDUs that are compiled to perform a SWAP where the original data atom is replaced with a ``fresh" reservoir atom in which the quantum information is now stored, and thus serves as the new data atom. As depicted in Fig.~\ref{fig:refilling}, these SWAP LDUs reduce the need for an extra refilling step after an LDU is performed, automatically supplying a re-initialized qubit in case of an atom loss.  In addition, this could also allow for effective cooling of the data carriers by replacement of atoms.  Heating incurred by, for example, repeated entangling operations or non-adiabatic trap movements may then be effectively mitigated mid-circuit without disturbing the quantum information.
In such a scheme, the fresh atoms would need to be swapped in sufficiently frequently to allow for high-fidelity operation of the entangling gates of the LDU before too much heat accumulates in any one atom.
These LDUs could also be efficiently incorporated into measurement-free quantum error correction protocols~\cite{omanakuttan2024faulttolerant}.
SWAP LDUs of this nature have the potential to be a powerful tool for reaching a paradigm where the quantum information for a calculation significantly outlives the physical lifetime of any single atom. 

In the near term, such a regime may be reached by loading significantly more reservoir atoms than necessary for a computation \cite{EndresAncillaReadout} and replacing the data-carrier atoms continuously. For longer-term applications, recent progress on continuous atom reloading strategies suggests a way to continuously replenish atoms in a quantum register for indefinite operation \cite{AtomComputing1000, BernienContinuousOperation, SchreckAtomLaserReview,Zeiher_continuous_reloading}.

These prospects assume that the atom-loss probability increases as the data atom undergoes logic operations and that the ancilla atom presence probability is high after the LDU.
While the SWAP LDU has the refilling advantages over the standard LDU, its failure rate increases with increasing ancilla atom-loss rate before the LDU. Even a single failed round of the SWAP LDU causes data information loss, since the original data atom is measured.
In contrast, the standard LDU does not have this problem as the data atom is preserved at the end. Furthermore, using a three-outcome measurement with the standard LDU, ancilla losses can be efficiently detected allowing their deleterious effects to be suppressed by postselection or subsequent attempts at the LDU.
In state-of-the-art experiments, the ancilla loading fraction is expected to approach unity such that ancilla losses are expected to be a minor problem. 
As long as the ancilla loss rate is small, this minor disadvantage of SWAP LDUs is expected to be far outweighed by the advantages of free refilling and cooling by replacement (see Appendix \ref{sec:Apdx_failure_rates} for details).

\subsection{Teleportation-based SWAP LDU}

In the SWAP LDU, we can combine the second entangling gate and the following measurement into a measurement feedback gate instead, as in a one-bit teleportation circuit~\cite{Chuang_compiling2000,Fowler_coping_teleportation}. This circuit has previously been used as an LRU that re-initializes a qubit in the computational subspace in the case of a leakage error, but cannot reveal if the leakage error occurred~\cite{ Fowler_coping_teleportation,Aliferis_Terhal_2007_LRU}. In a quantum error correction setting, LRUs replace a leakage error with a Pauli error (in contrast, LDUs replace leakage with erasure), which can then be corrected by further rounds of error correction.

We consider this same circuit but with a three-outcome measurement instead: a 0 outcome results in the conclusion that no local operation is needed and a 1 outcome results in a local Z operation to be applied on the new data atom for a successful SWAP. 
A ``neither" measurement outcome reveals an atom loss, which has already been rectified with the new data atom carrying a $\ket{0}$ state. Thus, a three-outcome measurement upgrades this LRU into an LDU, detecting the leakage error in addition to replacing it with a Pauli error. The main advantage of this ``Z-teleportation" version of the LDU over the previously discussed circuit is that it requires only one entangling gate instead of two at the cost of a measurement feedback local gate.

In a similar vein, our LLSD is not specific to the Zeeman-sublevel, so hyperfine leakage errors are not detected by this version of the LDU in our experiment.
Given the long measurement times for neutral atoms, this teleportation-based SWAP LDU is likely to be useful instead of the SWAP LDU with two entangling gates in cases where
the conditional $Z$ operator can be propagated into post-processing.

This teleportation-based SWAP LDU requires local single-qubit rotations, and as we do not currently have local measurement capabilities, we apply the measurement-controlled $Z$ gate in post processing. We test the performance of this circuit and find that state $\ket{0}$ is transferred correctly to the ancilla atom in 597/624 = 95.7(8)\% of experiments and the state $\ket{1}$ is transferred correctly in 965/1007 = 95.8(7)\% of shots in which the data and ancilla atoms are retained. As a baseline, when no LDU is performed, the data qubit reads out correctly as $\ket{0}$ in 98.4(3)\% of shots and is read out correctly as $\ket{1}$ in 98.3(3)\% of experiments after state preparation using the local $z$ rotations, implying that only 2-3\% of the error stems from the LDU itself for the computational basis states. 

\begin{figure}
    \centering
    \begin{subfigure}{0.48\textwidth}
    \caption{\raggedright}
    \label{fig:teleportation_LDU}
        $\begin{array}{c}
        \Qcircuit @C=1em @R=1em {
        \lstick{\ket{\psi}_d} & \qw & \ctrl{1} & \gate{H} & \meter \cwx[1] &  \\
        \lstick{\ket{0}_a} & \gate{H} & \ctrl{-1} & \gate{H} & \gate{Z} & \rstick{\ket{\psi}_a} \qw 
        }
        \end{array}$
    \end{subfigure}
    
    \begin{subfigure}{0.48\textwidth}
    \caption{\raggedright}
    \label{fig:teleportation_LDU_MMZZ}
        $\begin{array}{c}
        \Qcircuit @C=0.7em @R=1em {
        \lstick{\ket{\psi}_d} & \qw & \multigate{1}{R_{zz}(\frac{\pi}{2})R_{x}(\pi)} & \gate{R_{y}(-\frac{\pi}{2})} & \gate{R_{z}(-\frac{\pi}{2})} &  \meter \cwx[1] &  \\
        \lstick{\ket{0}_a} & \gate{R_{x}(\frac{\pi}{2})} & \ghost{R_{zz}(\frac{\pi}{2})R_{x}(\pi)} & \gate{R_{y}(-\frac{\pi}{2})} & \gate{R_{z}(-\frac{\pi}{2})} & \gate{Z} & \rstick{\ket{\psi}_a} \qw \\
        & & & & & 
        }
        \end{array}$
    \end{subfigure}
    
    \begin{subfigure}{0.48\textwidth}
        \centering
        \caption{\raggedright }
        \vspace{-12pt}
        \includegraphics[width=0.9\textwidth]{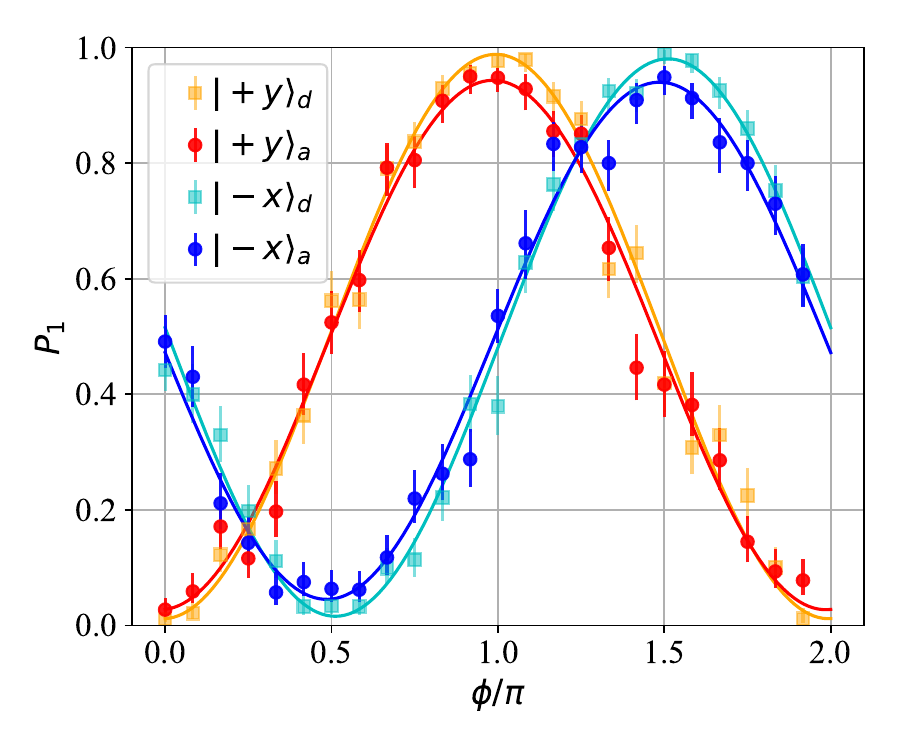}
    \end{subfigure}
    \caption{\raggedright Single-qubit teleportation LDU circuits compiled with typical gates \textbf{(a)} and our native gates \textbf{(b)}. 
    \textbf{(c)} Probability of measuring the $\ket{1}$ state as the phase ($\phi$) of a final $R_{\phi}(\frac{\pi}{2})$ pulse is scanned on the ancilla qubit ($a$) after teleportation of the $\ket{-x}$ and $\ket{+y}$ states (circles), with the same scan on the data qubit ($d$) with no LDU for reference (squares). Uncertainty markers on datapoints are 68\% Wilson score intervals.}
    \label{fig:teleport}
\end{figure}

Data atom loss errors are detected via LLSD in 929/7150 $\approx 13\%$ shots of the experiment (aggregating all input states).
In the case that the data atom is lost, we find that the remaining ancilla atom is measured to be in $\ket{0}$ in 396/794 $\approx 50\%$ shots, consistent with initialization of the ancilla in $\ket{-x}_a$ after the LDU in the case of detected leakage. We initialize in $\ket{-x}_a$ instead of $\ket{0}_a$ on the ancilla in the case of a leakage error because our circuit (Fig.~\ref{fig:teleportation_LDU_MMZZ}) differs from the teleportation LDU compiled with CZ and H (Fig.~\ref{fig:teleportation_LDU}) by a local rotation that does not impact the logic operation of the LDU.

More importantly, the original ancilla atom (new data atom) is similarly missing at the end of the computation in 889/7150 $\approx 12.4\%$ shots. 
Both atoms are lost in 135/7150 $\approx 1.9\%$ shots, consistent with independent random loss of the two atoms. While in this experiment, loss of the ancilla atom is mitigated by postselection on ancilla atom retention, (as made possible by LLSD) this loss reveals a fundamental limitation of the SWAP LDU atom replacement method. 
This effect is exaggerated here by a \textit{technical} limitation of high atom loss rate, specific to our system. 
This fundamental limit is set by the combined loading accuracy of the ancilla atom and its retention probability after the LDU. For state-of-the-art experiments, loading accuracy approaches unity \cite{Bluvstein2022, EndresBenchmarking} and loss probability after a single entangling gate can be suppressed to a conservative estimate of $\leq$ 0.005 \cite{Evered2023}. 
We note that initialization errors of this form have been studied in Ref.~\cite{BrownSurfaceCodeLeakage} in the context of embedding SWAP LRUs into a surface code, and appropriate scheduling of LRUs was shown to still produce a fault-tolerant implementation.

To check the coherence of the quantum information after the teleportation-based LDU, we teleport a $\ket{+y}$ and a $\ket{-x}$ state and perform a phase scan of an effectively-local final $R_{\phi}(\frac{\pi}{2})$ pulse to reveal the phase of the resulting superposition state on the ancilla atom; see Fig.~\ref{fig:teleport}. 
As the physical local rotations can only perform $R_z(\theta)$, we split the global
$R_{\phi}(\frac{\pi}{2})$ into two $R_{\phi}(\frac{\pi}{4})$ pulses and apply a local $R_z(\pi)$ on the original data qubit in-between to effectively cancel its participation in the $R_{\phi}$ gate  \cite{EndresLocalZ} \footnote{Note that a final, local $R_z(\pi)$ pulse is generally needed, but omitted here as it is followed by a measurement.}.
Compared to the direct data atom measurement in the case of no LDU, the contrast of the resulting teleported state is reduced from 96(1)\% to 90(2)\% for the $\ket{-x}$ state and from 98(1)\% to 92(2)\% for the $\ket{+y}$ state. 

\section{Summary and Outlook}

The experiments in this work have demonstrated proof-of-principle nondestructive detection of atom loss errors and other major leakage channels in neutral-atom quantum computing platforms via LDUs, converting leakage errors into erasure errors. Rydberg leakage is converted to atom loss via tweezer anti-trapping and we choose single- and two-qubit gates that are sensitive to Zeeman levels, which effectively makes hyperfine-leaked atoms transparent to the control lasers. The LDUs can then detect all three major leakage pathways (atom loss, Rydberg leakage, and hyperfine leakage) by imprinting the information of qubit presence/absence onto an ancilla atom.

We benchmarked the performance of these LDUs using our LLSD capability, which can detect atom loss projectively. We first explored the implementation of the standard LDU, which detects atom-loss errors with $\approx 93.4\%$ accuracy. In the event of ancilla loss before the LDU, this LDU might give the wrong outcome but it does not destroy the encoded information. Furthermore, we note that the three-outcome measurement capability makes the implementation of this LDU robust against ancilla losses.

Next, we considered the implementation of a SWAP-style LDU, which maps the qubit information onto a new ancilla atom. This LDU might have compiling advantages in a reconfigurable atom array, in the form of free refilling in case of atom-loss and cooling by replacement since hot data atoms may be swapped out for cold ancilla atoms. However, SWAP LDUs can destroy data-qubit information when ancilla atoms are lost before the LDU (unlike the standard LDU). Fresh ancilla losses are only a minor problem in state-of-the-art systems so we expect this disadvantage to be a minor problem when compared with the advantages of refilling.

We then considered a teleportation-based version of the SWAP LDU which only has one entangling gate followed by a measurement feedback gate. This works as an LDU due to our three-outcome measurement but can still work as an LRU for a two-outcome measurement. In a quantum error correction setting, this LRU would still be a sufficient gadget to convert leakage errors to Pauli errors but not into erasure errors. We implement this LDU with a $\approx95.7\%$ success rate, consistent with the reduced number of two-qubit gates.

The SWAP LDU is a promising approach to keeping a quantum register of atoms cold (and full) for long circuits without adding undue experimental burden. Similar to sympathetic cooling in trapped ions \cite{SympatheticCooling_Wineland_PRA_203}, 
alternate proposed approaches to cooling a quantum register of atoms without disrupting the internal state include immersion in a degenerate gas \cite{SpethmannSingleAtomInBEC, ZollerSympatheticCoolingAtoms}, scattering light in a coherence-preserving subspace \cite{Omanakuttan_future, omanakuttan2024faulttolerant}, cavity-assisted cooling \cite{ZollerCavityCooling}, and Rydberg-mediated phonon exchange \cite{GorshkovRydbergCooling}.  These approaches may require additional complexity such as laser cooling of additional atomic species, presence of specific level structure, or trapping forces for atoms with Rydberg character. By contrast, the SWAP LDU is done by design with a single atomic species and only circuit-based control already required for a quantum processor. The only experimental overhead incurred is in the loading of additional atoms in the system, either by loading extra atoms before the circuit or continuously loading in a spatially-separated region \cite{AtomComputing1000}. 
This extra atom loading requirement is well within reach of current neutral-atom quantum processors, as repeated readout of a data atom using multiple ancillas (via extra atoms loaded before the circuit) has already been demonstrated in \cite{EndresAncillaReadout}.

The majority of the errors in this work stem from the limited fidelity of entangling gates and local addressing operations in our apparatus. State-of-the-art neutral-atom systems \cite{Evered2023, EndresAncillaReadout, cao2024multiqubit, peper2024spectroscopymodeling171ybrydberg} could realistically already implement these LDUs with  $>97\%$ success rate, limited by the combination of two-qubit gate and SPAM errors. 
With current state-of-the-art minimum SPAM and high-fidelity entangling gate protocols \cite{PritchardBenchmarking}, LDUs could be performed with $\geq$ 99.6\% accuracy. 
This performance is near or beyond previously-calculated thresholds for various LRU embedding schemes in, e.g., the toric code \cite{suchara2015leakage}. More tailored schemes making full use of the erasure conversion of an LDU may have even higher thresholds \cite{Aliferis_Terhal_2007_LRU}.

The LDUs in this work are readily compiled using typical native gates of neutral-atom systems (e.g., CZ, H, X, Z) and can be composed using globally-addressed single-qubit rotations (as in Fig.~\ref{fig:globalstandard}), facilitating implementation on other experiments. 
Additionally, the standard LDUs used in this work were compiled for $I_d I_a$ in the case of no leakage and $X_a$ in the case of leakage simply for the convenience of circuit and result interpretation. The number of single-qubit gates may be reduced in some cases if we relax the requirements on the LDU to simply preserve the state information up to a local unitary and/or allow for flipped interpretation of the ancilla readout results. 
Similarly, hardware-optimized versions of these LDUs may be compiled to combine two entangling gates into a single gate of larger rotation angle to reduce the number of Rydberg laser pulses as in Fig.~\ref{fig:zzpi}.

These proof-of-principle experiments demonstrate the potential benefits of LDUs for neutral-atom quantum processors. 
The fact that the LDU circuits presented here are essentially the same as LRUs that have been embedded in error correcting codes~\cite{McEwen2021} shows great promise for the future implementation of the circuits studied here in such a fault-tolerant setting for neutral atoms. Future work includes analyzing the advantages of standard and SWAP LDUs under various strategies for insertion of LDUs, and their interplay with three-outcome measurements under realistic error models.

\begin{acknowledgments}
We thank Anupam Mitra for helpful discussions and feedback during the drafting of this manuscript. 
We thank Roger Ding for helpful discussions about sympathetic cooling. We thank Jacquilyn Weeks for assistance in technical writing. We also thank Andrew Landahl for helpful discussions on leakage errors in the context of error correction.

This work was supported by the Laboratory Directed Research and Development program at Sandia National Laboratories, a multimission laboratory managed and operated by National Technology and Engineering Solutions of Sandia LLC, a wholly owned subsidiary of Honeywell International Inc. for the U.S. Department of Energy’s National Nuclear Security Administration
under contract DE-NA0003525.
This written work is authored by an employee of NTESS. The employee, not NTESS, owns the right, title and interest in and to the written work and is responsible for its contents. Any subjective views or opinions that might be expressed in the written work do not necessarily represent the views of the U.S. Government. The publisher acknowledges that the U.S. Government retains a non-exclusive, paid-up, irrevocable, world-wide license to publish or reproduce the published form of this written work or allow others to do so, for U.S. Government purposes. The DOE will provide public access to results of federally sponsored research in accordance with the DOE Public Access Plan.
SAND2024-05904O

\end{acknowledgments}

\newpage

\appendix{}
\section{Supporting Data}

\subsection{Rydberg state anti-trapping measurement}
\begin{figure}[h]
    \centering
    \begin{subfigure}{0.48\textwidth}
        \includegraphics[width=0.9\textwidth]{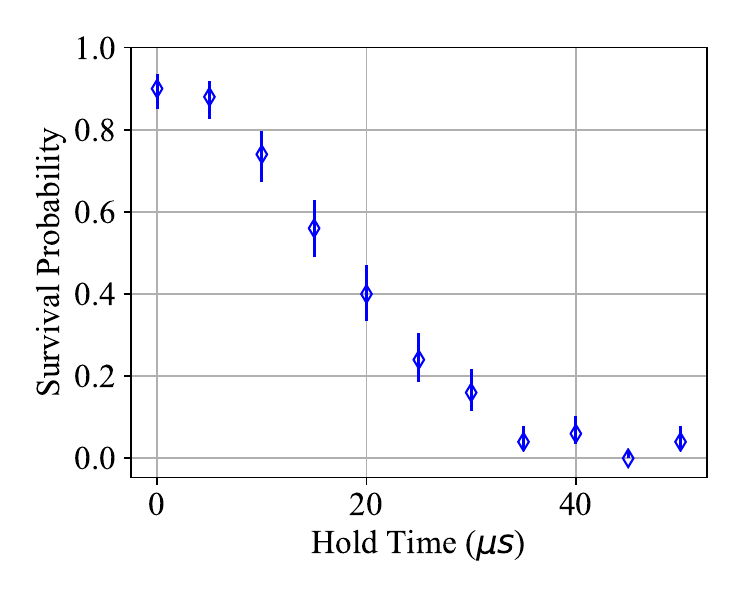}
    \end{subfigure}
    \caption{\raggedright Conversion of Rydberg state population errors to atom loss errors is done naturally by anti-trapping of atoms in the Rydberg state. Here we prepare the atom in the Rydberg state, hold the atom in the trap for a variable amount of time, and then de-excite the atom. Rydberg atoms are ejected from the trap after a few tens of $\mu$s. Uncertainty markers are 68\% Wilson score intervals.
    }
    \label{fig:antitrapping}
\end{figure}

Rydberg state anti-trapping speed measurement for the $64P_{3/2}$ state of Cs in our tweezers is measured by promoting a single atom to the Rydberg state, holding it in the trap for a variable amount of time, de-exciting it, and measuring the probability that the atom is retained in the trap. See Fig.~\ref{fig:antitrapping}.

\subsection{Full shot-by-shot data for standard LDU}
\label{sec:shotbyshot}

For completion and clarity, we include a logic table with the results of the standard LDU experiment in Table~\ref{tab:logic}.

Additionally, we provide further details on the difference between atom loss identification accuracy of the standard LDU in the case of intentionally prepared absent atoms and accidentally lost atoms. 
When we analyze the atom absent case of the standard LDU in only the instances where the data atom was intentionally prepared as absent, we find that the ancilla correctly reports the data atom as absent in 865/901 $\approx 96\%$ of shots, slightly better than the aggregate result of 919/984 $\approx 93.4\%$. This indicates that when the data atom is absent prior to the LDU, the resulting ancilla state accuracy is slightly better.
On the other hand, the reduction in accuracy overall (from contribution of the accidental losses) is indicative of atom loss errors during or after the LDU, which could allow for unknown propagation of errors after the atom loss event. In the 83 cases where the data atom was initially loaded, but lost before the final readout, the ancilla result reported the data atom as absent in 54 cases and present in 29 cases. 
These values are an indicator of when accidental atom loss is likely to occur in our system (before vs during or after the LDU).

\begin{table}[bh]
\resizebox{0.48\textwidth}{!}{
\begin{tabular}{|c|c|c|c|}
        \hline
        Data atom & Ancilla atom & Result & N \\
        \hline
        0 or 1 (present) & 0 & Correct present & 765 \\
        \hline
        0 or 1 (present)  & 1 & Incorrect absent & 49 \\
        \hline
        0 or 1 (present)  & Neither (absent) & Postselect against & 84 \\
        \hline
        Neither (absent) & 0 & Incorrect present & 65 \\
        \hline
        Neither (absent) & 1 & Correct absent & 919 \\
        \hline
        Neither (absent) & Neither (absent) & Postselect against & 118 \\
        \hline
\end{tabular}
}
\caption{\raggedright Logic table and results for standard LDU experiment. In a total of 1000 shots of the experiment each, we attempt to prepare the data atom as present or absent (ancilla atom always intended to be present). For each shot, the LLSD outcome for the data atom (first column) and ancilla atom (second column) determine the our interpretation of the result (third column). Total number of shots for each outcome (N) are tabulated in the fourth column.  }
\label{tab:logic}
\end{table}

\subsection{Entangling gate pulse sequence and unitary}

The entangling gate throughout this work is the inner three pulses of Ref.~\cite{Mitra2020, Martin2021}. When possible, we choose the phase ($\phi$) of the spin echo pulse (global single-qubit $\pi$-pulse) such that the circuit is first-order robust to pulse area errors in the single-qubit laser \cite{CPMG}.

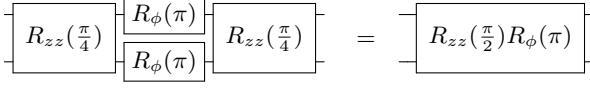
\begin{figure}
    \centering
        $\begin{array}{c}
        \Qcircuit @C=.35em @R=.35em {
           & & & & & & & \\
             & \multigate{1}{R_{zz}(\frac{\pi}{4})} 
             &  \gate{{R}_{\phi}(\pi)} 
             & \multigate{1}{R_{zz}(\frac{\pi}{4})} 
             & \qw 
             &  \push{\rule{1em}{0em}} 
             & \raisebox{-2.2em}{=}
             & \push{\rule{1em}{0em}} 
             & \qw 
             &  \multigate{1}{R_{zz}(\frac{\pi}{2})R_{\phi}(\pi)} 
             & \qw \\
            & \ghost{R_{zz}(\frac{\pi}{4})} 
            & \gate{{R}_{\phi}(\pi)} 
            & \ghost{R_{zz}(\frac{\pi}{4})}
            & \qw
            & \push{\rule{1em}{0em}}
            & 
            & \push{\rule{1em}{0em}} 
            & \qw
            & \ghost{R_{zz}(\frac{\pi}{2})R_{\phi}(\pi)}
            & \qw
            \\
            & & & & & & & 
        }
        \end{array}$
    \caption{\raggedright The entangling gate used in this work is composed of two UV ramp pulses ($R_{zz}(\frac{\theta}{2})$) with a global single-qubit $\pi$ pulse between, implementing the unitary $R_{zz}(\theta)R_{\phi}(\pi)$. Note that during the $R_{zz}(\frac{\pi}{4})$ pulses, additional uncalibrated single-particle phases are accumulated, but their contributions are removed with the echo pulse.}
    \label{fig:RzzCircuit}
\end{figure}

\subsection{Bell state fidelity measurement for entangling gate} \label{sec:Apdx_fidelity}

\begin{figure}[ht]
    \centering
    \begin{subfigure}{0.48\textwidth}
        \caption{\raggedright}
        \vspace{-12pt}
        \includegraphics[width=0.9\textwidth]{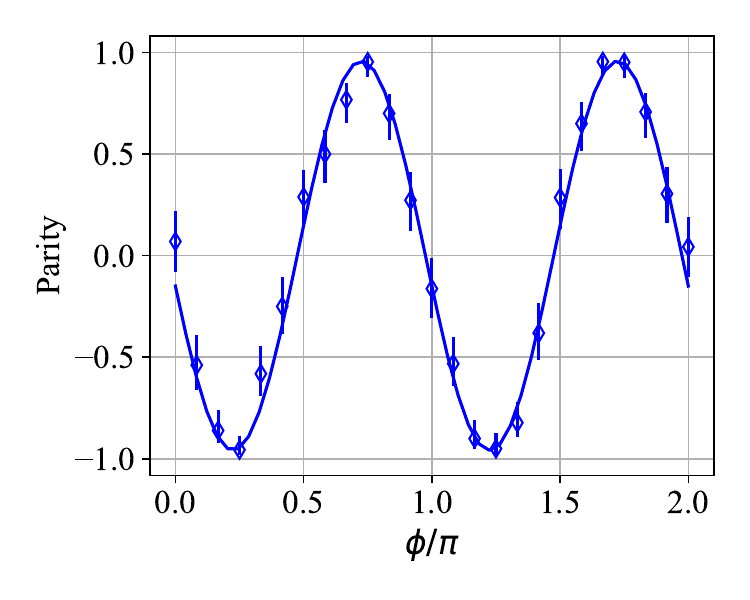}        
    \end{subfigure}
    
    \begin{subfigure}{0.48\textwidth}
        \caption{\raggedright}
        \vspace{-12pt}
        \includegraphics[width=0.9\textwidth]{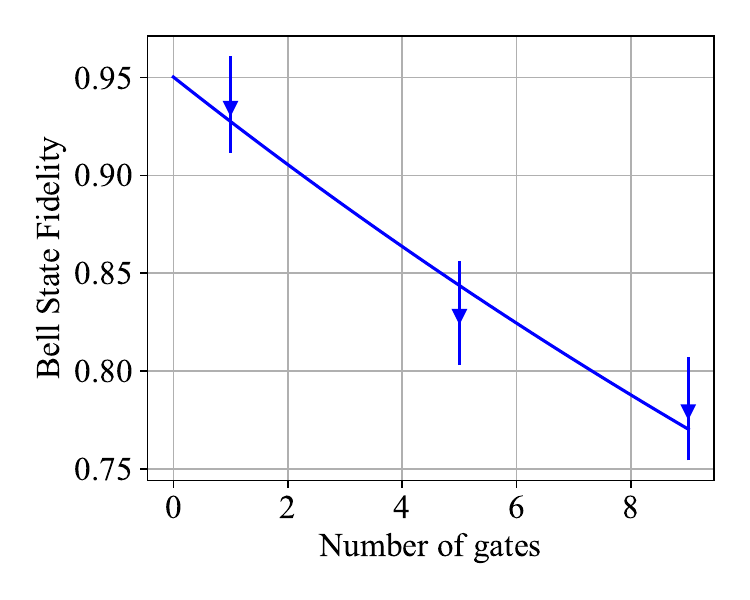}
    \end{subfigure}
    \caption{\raggedright Bell state fidelity measurement via parity scan of a final $R_{\phi}(\frac{\pi}{2})$ pulse and population measurements for 1, 5, and 9 of the $R_{zz}(\frac{\pi}{2})R_{y}(\pi)$ gate. 
    \textbf{(a)} Parity scan example data for 1 entangling gate applied. Displayed errorbars are 68\% Wilson score confidence intervals. 
    \textbf{(b)} Fidelity as a function of the number of loops is fit to an exponential with a decay constant of 30(5) loops. Errorbars are derived from 68\% Wilson score intervals on population measurements and fitting uncertainty on the parity contrast.}
    \label{fig:bellfidelity}
\end{figure}

Data used in determining the two-qubit gate fidelity reported in this work is included in Fig.~\ref{fig:bellfidelity}. We note that the fit to the parity oscillation uses Gaussian-distributed errors (Wald intervals), and may slightly overestimate contrast \cite{EndresAncillaReadout}. As discussed in the main text, these results include ``natural" postselection against Rydberg leakage errors via anti-trapping and postselection against atom loss errors with LLSD. Fidelity ($\mathcal{F}$) is assessed using population measurements and parity contrast according to \cite{Sackett2000, Kim2009, ManningThesis, Figgatt2019}

\begin{equation}
    \mathcal{F} = \frac{1}{2}\left(\rho_{00}+\rho_{11}\right) + \frac{1}{2}A ,
\end{equation}
where $A$ is the fitted parity contrast and $\rho_{ii}$ is the measured population in the $\ket{ii}$ state. 

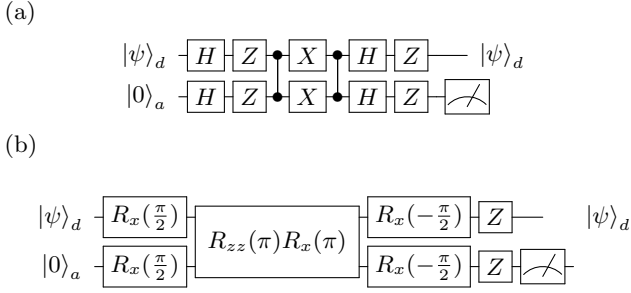
\begin{figure}[ht]
    \centering
    \begin{subfigure}{0.48\textwidth}
    \caption{\raggedright}
    \label{fig:globalstandard}
    $\begin{array}{c}
        \Qcircuit @C=.35em @R=.35em {
                & \lstick{\ket{\psi}_d} & \gate{H} & \gate{Z}
                & \ctrl{1} & \gate{X} & \ctrl{1} &  \gate{H} & \gate{Z} & \qw & \rstick{\ket{\psi}_d} \qw \\
                & \lstick{\ket{0}_a} & \gate{H} &  \gate{Z} 
                & \ctrl{-1} & \gate{X} & \ctrl{-1} & \gate{H} & \gate{Z} & \qw & \meter 
                }
    \end{array}$
    \end{subfigure}

    \begin{subfigure}{0.48\textwidth}
        \caption{\raggedright}
        \label{fig:zzpi}
        $\begin{array}{c}
        \Qcircuit @C=.35em @R=.35em {
            & & & & & \\
            & \lstick{\ket{\psi}_d} & \gate{{R}_{x}(\frac{\pi}{2})} & \multigate{1}{R_{zz}(\pi)R_x(\pi)} &  \gate{{R}_{x}(-\frac{\pi}{2})} & \gate{Z} & \qw & \rstick{\ket{\psi}_d}  \\
            & \lstick{\ket{0}_a} &  \gate{{R}_{x}(\frac{\pi}{2})} & \ghost{{R_{zz}(\pi)R_{x}(\pi)}} & \gate{{R}_{x}(-\frac{\pi}{2})} & \gate{Z} & \meter & \qw \\
            & & & & & 
        }
        \end{array}$
    \end{subfigure}

    \caption{\raggedright Alternate versions of the LDUs.
    \textbf{(a)} Global-addressing only version of the standard LDU, using gates native to typical neutral-atom hardware.
    \textbf{(b)} ``Hardware optimized" version of the standard LDU that combines two fully-entangling $R_{zz}(\frac{\pi}{2})R_{\phi}$ gates into a single $R_{zz}(\pi)R_{\phi}$ gate to reduce the number of Rydberg laser pulses. 
    }
    \label{fig:alternatecircuits}
\end{figure}

\subsection{LDU failure due to Ancilla loss} \label{sec:Apdx_failure_rates}

Here we consider the failure rates of the LDUs due to ancilla atom loss before the LDU, ignoring the errors caused by the gates in the LDUs and measurements. A standard LDU cannot detect data qubit leakage in the event of ancilla loss. With a three-outcome measurement, ancilla loss can be detected, which means even though the LDU fails, the failure itself can be detected. With a two outcome measurement, we cannot detect when data leakage happens, but this can be circumvented by multiple rounds of LDUs. In either scenario, while the single round of LDU fails, the data qubit is preserved in the data atom, making repeated measurements possible.

For the SWAP LDU, if the ancilla atom is missing, the entangling gates do not work. When the data atom is measured, it results in irretrievable loss of encoded information, in addition to a new leaked data qubit. This could be catastrophic if ancilla loss rates are high. Since fresh ancilla loss rates are quite low in state-of-the-art systems, we expect this to be insignificant. A thorough analysis of this trade off in a fault-tolerance setting would be necessary before large scale implementations. In addition, three outcome measurement can actively measure the original data atom loss while it is only inferred in a two-outcome measurement. This extra information could be useful in designing error correction decoders.

\subsection{Generic matrix for standard LDU}

Here we provide the process matrix for the standard LDU. The data atom may be in states $\left\{0, 1, \ell \right\}$, where $\ket{\ell}$ represents leakage out of the qubit subspace. In the case that the data atom has not leaked, the LDU performs the identity, and in the case that the data atom has leaked, the LDU performs a bit flip on the ancilla. An unimportant phase ($\phi$) may accompany the bit flip and has no logical consequence on the interpretation of the result. See Fig.~\ref{fig:alternatecircuits} for compilations of LDUs using global rotations for neutral atom specific gates.

\begin{equation}
    \label{eq:standardmatrix}
    \bordermatrix{ & \bra{00} & \bra{01} & \bra{10} & \bra{11} & \bra{\ell 0} & \bra{\ell 1} \cr
       \ket{00} & 1 & 0 & 0 & 0 &  &  \cr
       \ket{01} & 0 & 1 & 0 & 0 &  &  \cr
       \ket{10} & 0 & 0 & 1 & 0 &  &  \cr
       \ket{11} & 0 & 0 & 0 & 1 &  &  \cr
       \ket{\ell 0} &  &  &  &  & 0 & e^{i\phi} \cr
       \ket{\ell 1} &  &  &  &  & e^{-i\phi} & 0 \cr }
\end{equation}

\subsection{Low-loss state detection (LLSD)}
\label{sec:Apdx_LLSD}

Low-loss state detection is performed as in the fiber-coupled version of Ref.~\cite{ChowDetection}. Fluorescence from each trap site is individually fiber coupled (see Fig.~2.8 of Ref.~\cite{HankinThesis}) and delivered to detectors. 
In the first stage of detection, a probe laser beam resonant with the $6S_{1/2}, F=4$ to $6P_{3/2}, F'=5$ transition is turned on such that the $\ket{1}$ qubit state scatters photons in a closed cycling transition while the $\ket{0}$ qubit state remains dark. 
In the second stage, a repump beam (resonant with $6S_{1/2}, F=3$ to $6P_{3/2}, F'=4$) and 3D Doppler cooling beams are turned on alternating with the probe beam such that atoms that were initially in either qubit state are pumped into the bright manifold and scatter photons. During this stage, the detector is gated to be active only during the probe beam illumination times to avoid background scatter from the 3D cooling beams. 
In both stages, the bright and dark photon count distributions are separated by simple thresholding. If above threshold counts are detected in the first stage of detection, we infer the atom was in the $\ket{1}$ state. If above threshold counts are detected in only the second stage, we infer the atom was in the $\ket{0}$ state. If below threshold counts are detected in both stages, we infer the atom was absent. 

    


\newpage

\bibliography{main}





\end{document}